\documentclass[12pt]{article}
\usepackage[utf8]{inputenc}
\usepackage[english]{babel}
\usepackage{graphicx}
\usepackage{amsmath}
\usepackage{amsthm}
\usepackage{amsmath,amssymb} 
\usepackage{subcaption}
\usepackage[
    backend=biber,
    sorting=none
]{biblatex}
\usepackage{color}
\usepackage{pstricks}
\usepackage{pstricks-add}
\usepackage{pst-plot}
\usepackage{listings}
\usepackage[crop=off]{auto-pst-pdf}
\usepackage{oubraces}
\usepackage{cancel}
\usepackage[ruled,vlined]{algorithm2e}
\usepackage{comment}

\newtheorem{theorem}{Theorem}
\newtheorem{lemma}{Lemma}
\newtheorem{definition}{Definition}

\DeclareMathOperator{\north}{up}
\DeclareMathOperator{\east}{right}
\DeclareMathOperator{\northeast}{upright}
\DeclareMathOperator{\south}{down}
\DeclareMathOperator{\west}{left}
\DeclareMathOperator{\southwest}{downleft}
\DeclareMathOperator{\AndOp}{\textup{\textbf{and}}}
\DeclareMathOperator{\assign}{\longleftarrow}
\SetKw{KwAnd}{and}
\SetKw{KwBreak}{break}
\SetKw{KwGoto}{goto}
\SetKwFunction{FDelete}{delete}
\SetKwFunction{FTest}{test}
\SetKwFunction{FSort}{sort}
\SetKwFunction{FChunkSort}{chunk-sort}

\SetKwProg{KwFunction}{function}{}{end}

\newcommand{\elem}{{\circ}}
\newcommand{\xelem}{{\xcancel{\circ}}}

\newcommand{\smallGrid}{
    \psline(0, -0.5)(0, 3.5)
    \psline(1, -0.5)(1, 3.5)
    \psline(2, -0.5)(2, 3.5)
    \psline(3, -0.5)(3, 3.5)

    \psline(-0.5, 0)(3.5, 0)
    \psline(-0.5, 1)(3.5, 1)
    \psline(-0.5, 2)(3.5, 2)
    \psline(-0.5, 3)(3.5, 3)

    \psdots(0, 0)(1, 0)(2, 0)(3, 0)
    \psdots(0, 1)(1, 1)(2, 1)(3, 1)
    \psdots(0, 2)(1, 2)(2, 2)(3, 2)
    \psdots(0, 3)(1, 3)(2, 3)(3, 3)
}

\newcommand{\manyDots}[1]{
    \rput(0, #1){
        \psdots(0, 0)(1, 0)(2, 0)(3, 0)(4, 0)(5, 0)(6, 0)(7, 0)(8, 0)(9, 0)(10, 0)(11, 0)(12, 0)(13, 0)
    }
}

\newcommand{\bigDoubleGrid}{
    \psline(1, 0)(1, 6)
    \psline(2, 0)(2, 6)
    \psline(3, 0)(3, 6)
    \psline(4, 0)(4, 6)
    \psline(5, 0)(5, 6)
    \psline(6, 0)(6, 6)
    \psline(7, 0)(7, 6)
    \psline(8, 0)(8, 6)
    \psline(9, 0)(9, 6)
    \psline(10, 0)(10, 6)
    \psline(11, 0)(11, 6)
    \psline(12, 0)(12, 6)
    \psline(13, 0)(13, 6)

    \psline(0, 1)(13, 1)
    \psline(0, 2)(13, 2)
    \psline(0, 3)(13, 3)
    \psline(0, 4)(13, 4)
    \psline(0, 5)(13, 5)
    \psline(0, 6)(13, 6)
    
    \psline[linestyle=dashed](7, 7)(13, 7)
    \psline[linestyle=dashed](7, 6)(7, 7)
    \psline[linestyle=dashed](8, 6)(8, 7)
    \psline[linestyle=dashed](9, 6)(9, 7)
    \psline[linestyle=dashed](10, 6)(10, 7)
    \psline[linestyle=dashed](11, 6)(11, 7)
    \psline[linestyle=dashed](12, 6)(12, 7)
    \psline[linestyle=dashed](13, 6)(13, 7)
    \psdots(7, 7)(8, 7)(9, 7)(10, 7)(11, 7)(12, 7)(13, 7)

    \manyDots{0}
    \manyDots{1}
    \manyDots{2}
    \manyDots{3}
    \manyDots{4}
    \manyDots{5}
    \manyDots{6}

    \psline[arrows=->,linewidth=2pt](0, 0)(14, 0)
    \psline[arrows=->, linewidth=2pt](0, 0)(0, 7)
    \rput(14, 0.3){$i$}
    \rput(0.3, 7){$j$}
}

\providecommand{\keywords}[1]
{
  \textbf{\textit{Keywords:}} #1
}

\title{Discrete Frechet distance for closed curves}

\author{
  Evgeniy Vodolazskiy\\
  \texttt{waterlaz@gmail.com}
}

\begin{document}
\maketitle

\begin{abstract}
The paper presents a discrete variation of the Frechet distance between closed curves,
which can be seen as an approximation of the continuous measure.
A rather straightforward approach to compute the discrete Frechet distance between
two closed sequences of $m$ and $n$ points using binary search takes $\mathcal{O}(mn \log mn)$ time.
We present an algorithm that takes $\mathcal{O}(mn \log^* mn)$ time,
where $\log^*$ is the iterated logarithm.
\end{abstract}

\section{Introduction}
An intuitive definition of the discrete Frechet metric between closed curves involves two frogs connected with a string.
Suppose that there are two cycles of $m$ and $n$ stones respectively. 
The frogs may pick a starting stone (each frog from its respective cycle of stones).
Then, at each moment in time a frog can either stay put or jump to the next stone in the cycle.
The frogs can't skip stones and can't go back. 
The question is whether the frogs can complete exactly one cycle each and end up at their starting stones
without tearing the string.
The shortest length of the string that allows this is called the discrete Frechet distance between two cycles.

The Frechet metric as a tool to measure curve similarities has been introduced by Alt and Godau \cite{frechet}.
They have also provided a variation of the metric for closed curves, 
which is more appropriate for comparing image contours for example,
and proposed an algorithm that solves a decision problem of determining whether the Frechet 
distance between two closed polygonal curves is bigger than a given number in $\mathcal{O}(mn \log mn)$ time.
This result has been later improved by Schlesinger et al. \cite{schles2016} with 
an $\mathcal{O}(mn)$ time algorithm.

Computing the exact value of the Frechet distance between polygonal curves is known to
be a difficult problem even though Alt and Godau did provide an $\mathcal{O}(mn\log mn)$ algorithm
for open curves and an $\mathcal{O}(mn\log^2 mn)$ algorithm for closed curves \cite{frechet}.
This difficulty has led to the introduction of an approximation called the discrete Frechet distance \cite{eiter1994},
which can be computed for open curves in $\mathcal{O}(mn)$ time using dynamic programming.
This result has been later improved for the case of two dimensions by Pankaj K. Agarwal et al.
with a subquadratic algorithm \cite{agarwal2013}.
This paper introduces a variant of discrete Frechet distance between closed curves and proposes an
algorithm to compute it in $\mathcal{O}(mn \log^* mn)$ time, where $\log^*$ is the iterated logarithm.
$$
    \log^*(n) = 
        \begin{cases}
            0, & n \leq 1, \\
            1 + \log^*(\log n), & \text{otherwise}.
        \end{cases}
$$

\section{Problem formulation}

Consider a metric space $M$ with a metric $d:M \times M \to \mathbb{R}$.
Let $U = (u_1, u_2, u_3, \dots, u_m), u_i \in M$ and $V = (v_1, v_2, v_3, \dots, v_n), v_j \in M$ be
two sequences of $m$ and $n$ points respectively from the metric space $M$. 
A coupling of these two sequences is a sequence 
$$
    L = \big((u_{a_1}, v_{b_1}), (u_{a_1}, v_{b_1}), (u_{a_2}, v_{b_2}), \dots, (u_{a_l}, v_{b_l}) \big)
$$
of distinct pairs such that $a_1 = 1, b_1 = 1, a_l = m, b_l = n$ and 
either $a_{t+1} = a_t$ or $a_{t+1} = a_t+1$ and either $b_{t+1} = b_t$ or $b_{t+1} = b_t+1$ for all $t=\overline{1, l-1}$.
The length $||L||$ of coupling $L$ is the largest distance between pairs of points in the sequence:
$$
||L|| = \max_{t \in \{1, \dots, l\}} d(u_{a_l}, v_{b_l}).
$$
Denote the set of all couplings $C(U, V)$.
\begin{definition} \cite{eiter1994}
    Discrete Frechet distance between two sequences $U$ and $V$ is the minimal length of all couplings between $U$ and $V$
    $$
        \delta_{dF}(U, V) = \min_{L \in C(U, V)} ||L||.
    $$
\end{definition}
Naturally, a sequence of points $(u_1, u_2, u_3, \dots, u_m)$ has a starting point $u_1$ and an ending point $u_m$. 
One can say that for all integers $i, 2 \leq i \leq m-1,$ any point $u_i$ has 
a previous point $u_{i-1}$ and a next point $u_{i+1}$, 
but the point $u_1$ does not have a previous point and the point $u_m$ does not have a next point.
If, however, we define $u_m$ to be the previous point of $u_1$ and $u_1$ to be the next point of $u_m$,
we get a closed sequence.
\begin{definition}
    \sloppy A cyclic shift $S(U, s)$ of a sequence $U = (u_1, u_2, u_3, \dots, u_m)$ by an integer number $s \in \{0, 1, \dots, m\}$ is
    a sequence $U' = S(U, s) = (u'_1, u'_2, u'_3, \dots, u'_m)$ such 
    that $u'_i = u_{i+s}$ for $i + s \leq m$ and $u'_i = u_{i+s-m}$ for $i+s > m$.
\end{definition}
For a closed sequence any cyclic shift produces essentially the same sequence,
only the vertices are renumbered.
Discrete Frechet distance for closed sequences differs in that one does not know 
the starting points in both sequences (since there are no starting points).
\begin{definition}
\sloppy Discrete Frechet distance between closed sequences 
$U = (u_1, u_2, u_3, \dots, u_m)$ and $V = (v_1, v_2, v_3, \dots, v_n)$ is
$$
    \delta_{dcF}(U, V) = \min_{s_U \in \{0, \dots, m\}} \min_{s_V \in \{0, \dots, n\}}\delta_{dF}\big(S(U, s_U), S(V, s_V)\big).
$$
\end{definition}

\section{An $\mathcal{O}(mn \log mn)$ algorithm}
This section provides general concepts and ideas for computing the closed Frechet distance between sequences
that are used in Section~\ref{optimalSection} where the main result of the paper is given. 
To better illustrate the introduced concepts, this section ends with an algorithm 
that computes the closed Frechet distance in $\mathcal{O}(mn \log mn)$ time,
while the next section improves this result.

\emph{For closed sequences $U = (u_1, u_2, u_3, \dots, u_m)$ and $V = (v_1, v_2, v_3, \dots, v_n)$
the set of pairs of indices $D = \{1, \dots, m, m{+}1, \dots, 2m\} \times \{1, 2, \dots, n\}$ is
called the space \cite{frechet} of $U$ and $V$}. Notice that the points from $U$ are indexed twice.
Define $d:D \to \mathbb{R}$ the distance between a corresponding pair of points
$$
    d(i, j) = 
    \begin{cases}
        d(u_i, v_j), & i \leq m, \\
        d(u_{i-m}, v_j), & i > m.
    \end{cases}
$$
A point is allowed to move on the diagram in three directions: $\north$, $\east$ and $\northeast$ defined as follows:
$$
\north(i, j) = 
    \begin{cases}
        (i, j+1), & j < n, \\
        (i-m, 1), & j = n, i > m, 
    \end{cases}
$$
$$
\east(i, j) = 
        (i+1, j), \quad i < 2m, 
$$
$$
\northeast(i, j) = 
    \begin{cases}
        (i+1, j+1), & i < 2m,  j < n, \\
        (i-m+1, 1), & i \geq m, j = n.
    \end{cases}
$$
For most of the points on the diagram the definition of $\north$, $\east$ and $\northeast$ is straightforward.
The first exception is that sometimes the point on the edges can't move in one or more directions.
Whenever this happens, the corresponding value $\north(i, j)$, $\east(i, j)$ or $\northeast(i, j)$
is undefined.
The other exception is that whenever the point could end up at $(i, n+1), m+1 \leq i \leq 2m$ it is immediately moved
to the bottom of the diagram to $(i-m, 1)$. 
Even though points $(i, n+1), m+1 \leq i \leq 2m$ do not belong to the diagram, 
we will sometimes draw them on the diagram (see Figure~\ref{freespaceFig}) and consider $(i, n+1)$ to be identical to $(i-m, 1) \in D$.
\begin{figure*}[h!]
    \centering
    \begin{pspicture}(-0.75, -1)(14, 8)

        \bigDoubleGrid

        \rput(0, -0.4){$1$}
        \rput(1, -0.4){$2$}
        \rput(2, -0.4){$\dots$}
        \rput(3, -0.4){$i^*$}
        \rput(4, -0.4){$\dots$}
        \rput(5, -0.4){$m{-}1$}
        \rput(6, -0.4){$m$}

        \rput(7, -0.4){$m{+}1$}
        \rput(8.5, -0.4){$\dots$}
        \rput(10, -0.4){$m{+}i^*$}
        \rput(11, -0.4){$\dots$}
        \rput(12, -0.4){$2m{-}1$}
        \rput(13.2, -0.4){$2m$}

        \rput(-0.3, 0){$1$}
        \rput(-0.3, 1){$2$}
        \rput(-0.3, 2){$3$}
        \rput(-0.3, 3.5){$\dots$}
        \rput(-0.4, 5){$n{-}1$}
        \rput(-0.3, 6){$n$}

        \psline[showpoints=true, linewidth=3pt]
            (3, 0)(4, 1)(5, 2)(5, 3)
            (6, 4)(7, 5)(8, 5)(9, 6)(10, 7)
        \psline[arrows=->, linewidth=3pt]
            (3, 0)(4, 1)
        \psline[arrows=->, linewidth=3pt]
            (4, 1)(5, 2)
        \psline[arrows=->, linewidth=3pt]
            (5, 2)(5, 3)
        \psline[arrows=->, linewidth=3pt]
            (5, 3)(6, 4)
        \psline[arrows=->, linewidth=3pt]
            (6, 4)(7, 5)
        \psline[arrows=->, linewidth=3pt]
            (7, 5)(8, 5)
        \psline[arrows=->, linewidth=3pt]
            (8, 5)(9, 6)
        \psline[arrows=->, linewidth=3pt]
            (9, 6)(10, 7)
        \rput(11, 7.35){$(m{+}i^*, n{+}1) = (i^*, 1)$}

\end{pspicture}
    \caption{ The free space is the Cartesian product $\{1, \dots, 2m\} \times \{1, \dots, n\}$ of indices. 
              Any coupling between two cyclic sequences results in a monotone path from some point $(i^*, 1)$ 
              on the bottom of the free space that goes to $(m+i^*, n+1)$, which is virtually the same point $(i^*, 1)$. }
    \label{freespaceFig}
\end{figure*}

Consider a coupling between shifted sequence $U$ and shifted sequence $V$.
For any such coupling there is a corresponding monotone path on the diagram $D$ that starts
at some point $(i^*, 1), 1 \leq i^* \leq m,$ on the bottom of the diagram and goes to the top
to the point $(i^* + m, n+1) = (i^*, 1)$ completing a single full cycle on both sequences $U$ and $V$ (see Figure~\ref{freespaceFig}).
The length of a coupling is the maximal value $d(i, j)$ on the corresponding path.
Now, instead of finding a coupling with minimal length we can look for a monotone path that 
minimizes the maximal value $d(i, j)$ along its way.

The main idea of the presented algorithm is that it sorts all points on the diagram $D$ by value $d(i, j)$
in descending order. Then the algorithm goes through the sorted array making points from it forbidden and
checks whether there exists a monotone path that does not go through forbidden points.
We will show that this can be done efficiently.

\begin{figure*}[h!]
    \centering
    \newcommand{\smallGridIndexes}{
    \rput(1, -0.7){$i$}
    \rput(2, -0.7){$i{+}1$}
    \rput(-0.7, 1){$j$}
    \rput(-0.9, 2){$j{+}1$}
}

\begin{pspicture}(-1.3, -1)(10.5, 4)
        \smallGrid
        \smallGridIndexes

        \psdots[dotstyle=x, dotsize=8pt](1, 2)(2, 1)(2, 2)
        \rput(4.4, 1.5){\scalebox{1.5}{$\Rightarrow$}}
        \rput(6, 0){
            \smallGrid
            \smallGridIndexes
            \psdots[dotstyle=x, dotsize=8pt](1, 2)(2, 1)(2, 2)(1, 1)
        }
\end{pspicture}
    \caption{ When points $(i{+}1, j),\; (i, j{+}1),\; (i{+}1,j{+1})$ are forbidden,
        function ``test$(i, j)$'' (see Algorithm~\ref{algTestDelete}) makes $(i, j)$ forbidden as well 
        since no monotone path can go through $(i, j)$.}
    \label{testFigure}
\end{figure*}
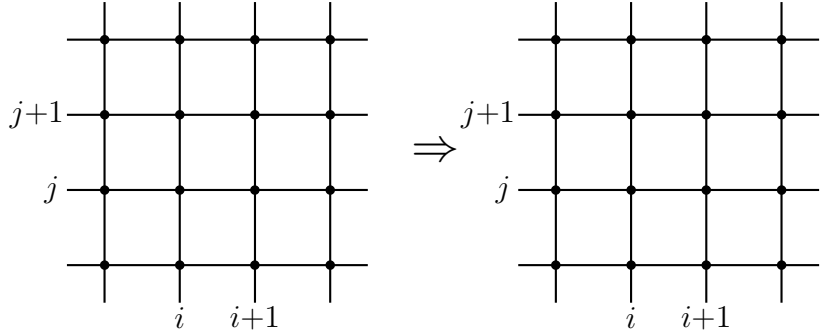

The algorithm starts with an array $q:D \to \{0, 1\}$ initialized to $q(i, j) = 1$ for all $(i, j) \in D$.
Value $q(i, j) = 0$ means that the point $(i, j)$ is forbidden and $q(i, j) = 1$ means that $(i, j)$ is allowed.
A counter $c$ initialized to $c=2mn$ is decreased by $1$ every time a point is made forbidden. 
This allows us to efficiently check whether there are any allowed points left by checking the condition $c>0$.
The algorithm constantly keeps track that there are no dead ends on the diagram. 
That is from any allowed point one can continue some monotone path to another allowed point.
When this is not possible, any point that is a dead end is also considered forbidden (see Figure~\ref{testFigure}).
This is done by calling ``test'' for the points in question.
\begin{figure*}[h!]
    \centering
    \begin{pspicture}(-1.3, -1)(5.5, 4)
        \smallGrid
        \rput(1, -0.7){$i{-}1$}
        \rput(2, -0.7){$i$}
        \rput(-0.9, 1){$j{-}1$}
        \rput(-0.7, 2){$j$}

        \psdots[dotstyle=x, dotsize=8pt](2, 2)
        \rput(0.85, 1.8){?}
        \rput(1.85, 0.8){?}
        \rput(0.85, 0.8){?}
\end{pspicture}
    \caption{ Function ``delete$(i, j)$'' makes the point $(i, j)$ forbidden and 
        calls ``test'' (see Algorithm~\ref{algTestDelete}) for 
        $(i{-}1, j),\; (i, j{-}1),\; (i{-}1,j{-1})$. Since $(i, j)$ became forbidden, 
        it may be that there are no more monotone paths through some of these three points.
    }
    \label{deleteFigure}
\end{figure*}
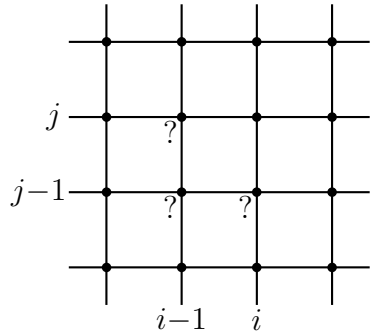
If some point $(i, j)$ is a part of a monotone path it can be reached from one of three directions:
$\west(i, j) = \east^{-1}(i, j)$, $\south(i, j) = \north^{-1}(i, j)$ and $\southwest(i, j) = \northeast^{-1}(i, j)$.
So to make sure that there are no dead ends whenever a point is made forbidden the three points that
are to the left, down or down-left are checked (see Figure~\ref{deleteFigure}).

\begin{algorithm}     
    \DontPrintSemicolon
    \KwFunction{\FDelete$(i, j)$}{
        $q(i, j) \assign 0$ \;
        $c \assign c - 1$\;
        \FTest$\big(\west(i, j)  \big)$\;
        \FTest$\big(\south(i, j)  \big)$\;
        \FTest$\big(\southwest(i, j)  \big)$\;
    }
    \KwFunction{\FTest$(i, j)$}{
        \If{$($\;$\quad\quad\begin{aligned} 
                q\big(i, j\big)=1 \; &\AndOp \; q\big(\north(i,j)\big)=0 \\
                               \,&\AndOp\;q  \big(\east(i,j)\big)=0 \\
                               \,&\AndOp\;q\big(\northeast(i,j)\big)=0
            \end{aligned}$\;$)$
            }{
                \FDelete$(i, j)$\;
            }
    }
    \caption{Function ``delete''}\label{algTestDelete}
\end{algorithm}

Algorithm~\ref{algTestDelete} provides function ``delete'', 
which allows one to start with a diagram where all points are allowed $q(i, j) = 1$,
then sequentially delete (make them forbidden) some points on the diagram and test whether there still is a monotone path.
Whenever a point is deleted, its neighbors to the left, down and down-left are checked with function ``test''.
This ensures that after an arbitrary number of calls to ``delete'' from any allowed point $(i, j)$ one can move 
either up, right or up-right to another allowed point.

Obviously, when all points on the diagram are forbidden there are no monotone paths on the diagram.
Less obvious, but no less true is that when there are still points that are not forbidden, 
a monotone path does exist as the following lemma states.
\begin{lemma}\label{pathExistsLemma}
Let some algorithm initialize $q(i, j) = 1$ for all $(i, j) \in D$ and $c = 2mn$.
If after a number of calls to function ``delete'' there are some allowed points ($c>0$)
then there exists a monotone path from a point $(i, 1)$ for some $i$ 
to a point $(i+m, n+1)$ that does not go through forbidden points.
\end{lemma}
\begin{proof}
\begin{figure*}[h!]
    \centering
    \begin{pspicture}(-0.75, -1)(14, 8)

    \bigDoubleGrid

    \psline[showpoints=true, linewidth=3pt]
        (1, 0)(2, 1)(3, 2)(3, 3)
        (4, 4)(5, 5)(6, 5)(7, 5)(8, 6)(9, 6)(10, 7)
    \psline[arrows=->, linewidth=3pt]
        (1, 0)(2, 1)
    \psline[arrows=->, linewidth=3pt]
        (2, 1)(3, 2)
    \psline[arrows=->, linewidth=3pt]
        (3, 2)(3, 3)
    \psline[arrows=->, linewidth=3pt]
        (3, 3)(4, 4)
    \psline[arrows=->, linewidth=3pt]
        (4, 4)(5, 5)
    \psline[arrows=->, linewidth=3pt]
        (5, 5)(6, 5)
    \psline[arrows=->, linewidth=3pt]
        (6, 5)(7, 5)
    \psline[arrows=->, linewidth=3pt]
        (7, 5)(8, 6)
    \psline[arrows=->, linewidth=3pt]
        (8, 6)(9, 6)
    \psline[arrows=->, linewidth=3pt]
        (9, 6)(10, 7)

    \psline[showpoints=true, linewidth=3pt]
        (3, 0)(4, 1)(5, 2)(6, 3)(7, 3)(8, 3)(9, 3)(10, 3)(11, 4)(12, 5)(12, 6)(12, 7)
    \psline[arrows=->, linewidth=3pt]
        (3, 0)(4, 1)
    \psline[arrows=->, linewidth=3pt]
        (4, 1)(5, 2)
    \psline[arrows=->, linewidth=3pt]
        (5, 2)(6, 3)
    \psline[arrows=->, linewidth=3pt]
        (6, 3)(7, 3)
    \psline[arrows=->, linewidth=3pt]
        (7, 3)(8, 3)
    \psline[arrows=->, linewidth=3pt]
        (8, 3)(9, 3)
    \psline[arrows=->, linewidth=3pt]
        (9, 3)(10, 3)
    \psline[arrows=->, linewidth=3pt]
        (10, 3)(11, 4)
    \psline[arrows=->, linewidth=3pt]
        (11, 4)(12, 5)
    \psline[arrows=->, linewidth=3pt]
        (12, 5)(12, 6)
    \psline[arrows=->, linewidth=3pt]
        (12, 6)(12, 7)

    \psdots[linewidth=3pt] 
        (5, 0)(6, 1)(7, 1)(8, 2)(9, 3)(9, 4)(10, 5)(11, 6)(11, 7)
    \psline[arrows=->, linewidth=3pt, linestyle=dashed]
        (5, 0)(6, 1)
    \psline[arrows=->, linewidth=3pt, linestyle=dashed]
        (6, 1)(7, 1)
    \psline[arrows=->, linewidth=3pt, linestyle=dashed]
        (7, 1)(8, 2)
    \psline[arrows=->, linewidth=3pt, linestyle=dashed]
        (8, 2)(9, 3)
    \psline[arrows=->, linewidth=3pt, linestyle=dashed]
        (9, 3)(9, 4)
    \psline[arrows=->, linewidth=3pt, linestyle=dashed]
        (9, 4)(10, 5)
    \psline[arrows=->, linewidth=3pt, linestyle=dashed]
        (10, 5)(11, 6)
    \psline[arrows=->, linewidth=3pt, linestyle=dashed]
        (11, 6)(11, 7)
    
    \rput(1, -0.4){$i_0$}
    \rput(3, -0.4){$i_1$}
    \rput(5, -0.4){$i_2$}
    
    \rput(9.8, 7.4){$m{+}i_1$}
    \rput(12.2, 7.4){$m{+}i_2$}
    \rput(11, 7.4){$m{+}i_3$}


\end{pspicture}
    \caption{ Eventually two consecutive paths will intersect. 
        Here the path from $(i_1, 1)$ to $(m{+}i_2, n+1)$ intersects the path from $(i_2, 1)$ to $(m{+}i_3, n+1)$.
        Which means that there is a path from $(i_2, 1)$ to $(m{+}i_2, n+1)$.}
    \label{pathsIntersectFig}
\end{figure*}
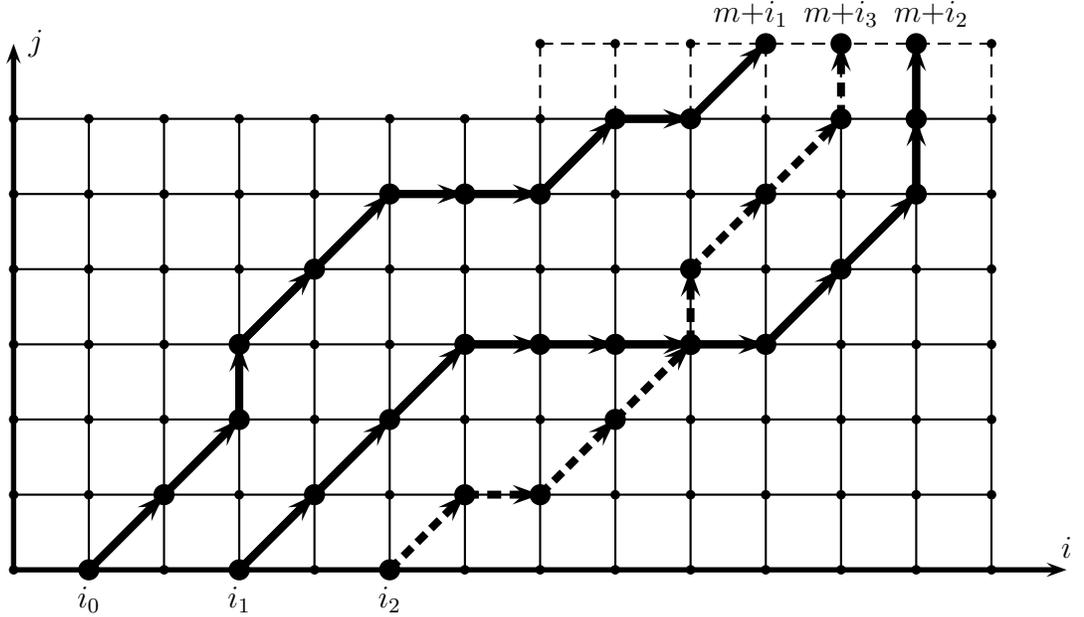
Indeed, since not all points are forbidden, there must be at least one allowed point $(i_0, 1), 1 \leq i_0 \leq m,$ 
on the bottom of the diagram. Otherwise, this would eventually lead to all points from $D$ to be forbidden.
Since $(i_0, 1)$ is allowed, one can sequentially move from it to another allowed point $(m{+}i_1, n{+}1) = (i_1, 1)$, 
which goes back to the bottom of the diagram and so on (see Figure~\ref{pathsIntersectFig}).
This means that there is an infinite sequence of paths that start from 
$$(i_0, 1), (i_1, 1), (i_2, 1), \dots$$
and go to 
$$(m{+}i_1, n{+}1), (m{+}i_2, n{+}1), (m{+}i_3, n{+}1), \dots.$$ 
The sequence $i_0, i_1, i_2, \dots$ can not be strictly monotone, 
i.e. neither $i_0 < i_1 < i_2 < \dots$ nor $i_0 > i_1 > i_2 > \dots$ is possible.
This means that eventually some path from $(i_t, 1)$ to $(m{+}i_{t+1}, n{+}1)$ will intersect 
a path from $(i_{t+1}, 1)$ to $(m{+}i_{t+2}, n{+}1)$. 
Therefore, there is a monotone path from $(i_{t+1}, 1)$ to $(m{+}i_{t+1}, n{+}1)$.
\end{proof}

We now can formulate the algorithm to compute the discrete Frechet distance.

\begin{algorithm}
\DontPrintSemicolon
$c \assign 2mn$ \;
\For{$(i, j) \in D$}{
    $q(i, j) \assign 1$\;
}
$A \assign \big\{\big(d(i, j), i, j\big) | (i, j) \in D \big\}$\;
\FSort{$A$} {\it // sort A in descending order by first component}\;
\For{$(d, i, j) \in A$}{
    \If{$(q(i, j) = 1)$}{
        \FDelete{$i, j$}\;
    }
    \If{$(c = 0)$}{
        \KwRet{$d$}\;
    }
}
\caption{Straightforward discrete Frechet distance}\label{mnlogmnAlg}
\end{algorithm}

\begin{theorem}\label{mnlogmnTheorem}
    Algorithm~\ref{mnlogmnAlg} takes $\mathcal{O}(mn \log mn)$ time.
\end{theorem}
\begin{proof}
The most time-consuming part of the algorithm is sorting the array $A$, 
which takes $\mathcal{O}(mn \log mn)$ time.
Indeed, function ``delete'' is called exactly $2mn$ times (including calls from ``test'').
And each call to ``delete'' results in no more than $3$ calls to ``test''.
Therefore, the rest of the algorithm after sorting takes $\mathcal{O}(mn)$ time and
the whole algorithm takes $\mathcal{O}(mn \log mn)$ time.
\end{proof}

\section{Improving asymptotic time}\label{optimalSection}
One can see that the most expensive part of Algorithm~\ref{mnlogmnAlg} is sorting the array.
The array, however, does not have to be fully sorted. 
Suppose that after sequentially calling ``delete'' the algorithm has made all points of $D$ forbidden
by eventually calling delete$(i, j)$ for element $a_t = (d, i, j)$ from $A$.
If the array is such that $a_t \leq a_i$ for $i<t$ and $a_t \geq a_i$ for $i>t$,
then $d$ is the discrete Frechet distance even though the array $A$ may not have been fully sorted.
This is an immediate consequence of the following lemma.
\begin{lemma}\label{LRLemma}
    Let some permutation of the array $A = \big\{\big(d(i, j), i, j\big) | (i, j) \in D \big\}$
    be divided into two subarrays $L \geq R$ such that all elements from $L$ are not less than any element from $R$.
    Let some algorithm initialize $q(i, j) = 1$ for all $(i, j) \in D$ and $c = 2mn$,
    and then call function ``delete'' for all elements from $L$.
    \[
        \underbrace{ \xelem \xelem \xelem \xelem \xelem \dots \xelem \xelem \xelem }_{L} \geq
        \underbrace{ \elem \elem \elem \elem \elem \dots \elem \elem }_{R}.
    \]
    Then the following statements are true:
    \begin{enumerate}
        \item
            If not all points from $D$ are forbidden ($c>0$) then 
            the subarray $R$ contains the value of closed discrete Frechet distance.
        \item 
            If all points from $D$ are forbidden ($c=0$) then
            the subarray $L$ contains the value of closed discrete Frechet distance.
    \end{enumerate}
\end{lemma}
\begin{proof}
    \begin{enumerate}
      \item
        According to Lemma~\ref{pathExistsLemma} if $c>0$ then there exists a monotone path 
        that goes through allowed points, which are points from $R$.
        The discrete Frechet distance is one of the values on a monotone path.
        Therefore, the maximum value along any path, where all points belong to $R$,
        is not greater than the maximum value along any path that passes through points from $L$.
        Therefore, the distance is one of the values from $R$.
      \item
        Suppose that the Frechet distance is not one of the values from $L$. 
        This means that there exists a monotone path 
        with all values $d(i, j)$ along its way that are strictly less than any value from $L$.
        Such a path can only go through points from $R$. 
        However, since $c=0$, there is no monotone path in $R$.
        Therefore, the Frechet distance is one of the values from $L$.
    \end{enumerate}
\end{proof}
We exploit this fact for the asymptotically better algorithm.

\begin{definition} \label{kChunkSortDef}
For an array $A =a_1a_2a_3\dots a_n$ of $n$ elements 
a $k$-chunk-sort is such a division of the array into $m$ arrays 
$A_1=a^1_1a^1_2a^1_3\dots a^1_{n_1},\; A_2=a^2_1a^2_2a^2_3\dots a^2_{n_2}, \; \dots, \; A_m=a^m_1a^m_2a^m_3\dots a^1_{n_m}$ that:
\begin{itemize}
    \item The concatenation of the arrays $A_1A_2\dots A_m$ is a permutation of the original array $A$.
    \item For any two arrays $A_i$ and $A_j$ such that $i<j$ any element from $A_i$ is not less than any element from $A_j$:
        $$i<j, a \in A_i, b \in A_j \Rightarrow a \geq b.$$
    \item The size $n_i$ of each array $A_i, i=\overline{1,m},$ does not exceed $\lceil n/k \rceil$:
        $$n_i = |A_i| \leq \lceil n/k \rceil.$$
\end{itemize}
The ordering of elements inside each array $A_i$ is irrelevant.
\end{definition}
The $k$-chunk-sorted array can be illustrated as follows:
$$
\underbrace{a^1_1a^1_2 \dots a^1_{n_1}}_{|A_1| \leq \lceil n/k \rceil} \geq 
\underbrace{a^2_1a^2_2 \dots a^2_{n_2}}_{|A_2| \leq \lceil n/k \rceil} \geq 
\underbrace{a^3_1a^3_2 \dots a^3_{n_3}}_{|A_3| \leq \lceil n/k \rceil} \geq \dots \geq 
\underbrace{a^m_1a^m_2 \dots a^m_{n_m}}_{|A_m| \leq \lceil n/k \rceil},
$$
where the $\geq$ symbol denotes that all elements to the left of $\geq$ are not less than the ones to the right of $\geq$.
Ideally, a $k$-chunk-sort would divide an array of $n$ elements into $k$ equal arrays of $n/k$ elements each.
This is not necessary though. 
Also note that an $n$-chunk-sort of $n$ elements is just a regular descending sort 
and a $1$-chunk-sort of an array is any permutation of the array. 

A straightforward practically efficient way to implement $k$-chunk-sort would be to use quicksort \cite{hoare1961} that stops
sorting when it reaches subarrays of size not greater than $\lceil n/k \rceil$.
For the sake of simplifying the analysis, however, 
the proof of the following theorem provides a less efficient and somewhat redundant yet 
asymptotically optimal algorithm.
\begin{theorem}
A $k$-chunk-sort of an array of $n$ elements can be done in $\mathcal{O}(n \log k)$ time.
\end{theorem}
\begin{proof}
We prove the theorem by doing a divide and conquer approach to the problem.
Given an array $A$ divide it into two arrays $A_1$ and $A_2$ with almost equal lengths $|A_1| = |A_2| \pm 1$
using a median $\mu$ of array $A$ as a pivot point. 
That is if $\mu$ is the median element of $A$, then $a_1 \geq \mu, a_1 \in A_1$ and $a_2 \leq \mu, a_2 \in A_2$.
This performs $2$-chunk-sort of $A$. 
To obtain $k$-chunk-sort we recursively call $\lceil k/2 \rceil$-chunk-sort on both $A_1$ and $A_2$
($\lceil\cdot\rceil$ denotes rounding the fraction up).
The median $\mu$ can be found in $\mathcal{O}(n)$ time \cite{blum1973} and therefore the $2$-chunk-sort
can be done in $\mathcal{O}(n)$ time. 
With each recursion depth the number of chunks doubles and the sizes of the subarrays halve.
Therefore, each depth of recursion takes $\mathcal{O}(n)$ time in total for all subarrays
and the maximal depth of recursion is $\mathcal{O}(\log k)$.
This gives an $\mathcal{O}(n \log k)$ algorithm.
\end{proof}

\begin{algorithm}
    \DontPrintSemicolon
    $k \assign 2$\;
    $L \assign \{\}$\;
    $R \assign \{\}$\;
    $B \assign \big\{\big(d(i, j), i, j\big) | (i, j) \in D \big\}$\;
    \While{$(k<2mn)$}{
        $c \assign 2mn$ \;
        \For{$(i, j) \in D$}{
\nl         $q(i, j) \assign 1$\; \label{initQLine}
        }
\nl     $B_1,B_2,\dots,B_l \assign $ \FChunkSort{$k, B$} \; \label{kChunkSortLine}
        $B_0 \assign L$\;
        \For{$s \assign 0$ \KwTo $l$}{
             \For{$(d, i, j) \in B_s$}{
                 \If{$(q(i, j) = 1)$}{
\nl                 \FDelete{$i, j$}\; \label{deleteLine}
                 }
\nl              \If{$(c = 0)$}{ \label{emptyDiagramLine}
                     \If{$(|B_s| = 1)$}{
                         \KwRet $d$ \;
                     } 
\nl                  $L \assign LB_1B_2\dots B_{s-1}$\; \label{newLLine}
\nl                  $R \assign B_{s+1}\dots B_l$\;     \label{newRLine}
\nl                  $B \assign B_s$\;                  \label{newBLine}
                     \KwGoto nextEpoch\;
                 }
             }
        }
        nextEpoch:\;
\nl     $k \assign 2^k$\;  \label{kGrowthLine}
    }
    \caption{Computing discrete Frechet distance}\label{logstarAlg}
\end{algorithm}

The proposed Algorithm~\ref{logstarAlg} runs in several epochs numbered $t$.
Each epoch starts with some permutation of the array $A = \big\{\big(d(i, j), i, j\big) | (i, j) \in D \big\}$ 
divided into three subarrays $L$, $B$ and $R$ such that elements from $L$ are not less than elements from $B$ 
and elements from $B$ are not less than elements from $R$
$$
\underbrace{ \elem \elem \elem \dots \elem }_{L} \geq
\underbrace{ \elem \elem \elem \elem \elem \dots \elem \elem \elem \elem \elem \elem \elem \elem \elem }_{B} \geq
\underbrace{ \elem \elem \elem \dots \elem \elem }_{R},
$$
where $\elem$ denotes one element from an array. 
Moreover, it is known that one of the elements from $B$ contains the value of closed discrete Frechet distance.
The algorithm performs a $k_t$-chunk-sort of the subarray $B$ for some $k_t$ defined later,
which divides elements from $B$ into $l$ arrays $B_1, B_2, \dots, B_l$ (line~\ref{kChunkSortLine}).
\[
\rlap{$
\underbrace{ \elem \elem \elem \dots \elem }_{L} \geq
\underbrace{ \elem \elem \elem }_{B_1} \geq
\underbrace{ \elem \elem \elem }_{B_2} \geq
\underbrace{ \elem \elem \elem }_{B_3} \geq
\underbrace{ \elem \elem \elem \elem }_{B_4} \geq \dots \geq
\underbrace{ \elem \elem \elem }_{B_l} \geq 
\underbrace{ \elem \elem \elem \dots \elem \elem }_{R}.
$}
\phantom{\elem \elem \elem \dots \elem \geq}
\overbrace{
    \phantom{
    \elem \elem \elem  \geq
    \elem \elem \elem \geq
    \elem \elem \elem  \geq
    \elem \elem \elem \elem  \geq \dots \geq
    \elem \elem \elem \geq }}^{k_t\text{-chunk-sort of } B}  
    \phantom{\elem \elem \elem \dots \elem \elem.}
\]
Each epoch starts with all of the diagram $D$ initialized as allowed (line~ \ref{initQLine}).
Then, the algorithm sequentially calls ``delete'' (line~\ref{deleteLine}) for the elements starting with $L$, $B_1$, $B_2$ and so on
until one of the calls to ``delete'' on an element of some subarray $B_s$ makes all of diagram $D$ forbidden 
($c=0$, line~\ref{emptyDiagramLine})
\[
\rlap{$
\underbrace{ \xelem \xelem \xelem \dots \xelem }_{L} \geq
\underbrace{ \xelem \xelem \xelem }_{B_1} \geq
\underbrace{ \xelem \xelem \xelem }_{B_2} \geq
\dots \geq
\underbrace{ \xelem \xelem \elem \elem }_{B_s} \geq \dots \geq
\underbrace{ \elem \elem \elem }_{B_l} \geq
\underbrace{ \elem \elem \elem \dots \elem \elem }_{R},
$}
\phantom{\elem \elem \elem \dots \elem \geq}
\overbrace{
    \phantom{
    \elem \elem \elem  \geq
    \elem \elem \elem \geq
    \dots \geq
    \elem \elem \elem \elem  \geq \dots \geq
    \elem \elem \elem \geq }}^{k_t\text{-chunk-sort of } B}
    \phantom{\elem \elem \elem \dots \elem \elem,}
\]
where $\xelem$ denotes elements on which the function ``delete'' has been called.
We can guarantee that $B_s$ contains the value of closed discrete Frechet distance (as a consequence of Lemma~\ref{LRLemma}) and
go to the next epoch $t+1$ with 
$L^{t+1} = LB_1B_2\dots B_{s-1},\;$
$B^{t+1} = B_s$ and
$R^{t+1} = B_{s+1}B_{s+2}\dots B_lR$ (lines~\ref{newLLine},~\ref{newRLine},~\ref{newBLine})
\[
\overunderbraces{
&\br{7}{L^{t+1}}
& &\br{1}{B^{t+1}}
& &\br{6}{R^{t+1}}
}{
&\elem \elem \dots \elem&  \geq
&\elem \elem \elem&  \geq
&\elem \elem \elem&  \geq \dots \geq
&\elem \elem \elem&  \geq
&\elem \elem \elem \elem& \geq 
&\elem \elem \elem&  \geq \dots \geq
&\elem \elem \elem&  \geq 
&\elem \elem \elem \dots \elem& \; &. 
}{
    &\br{1}{L} & &\br{1}{B_1} & &\br{1}{B_2} & & \br{1}{B_{s-1}}
    & & \br{1}{B_s}& & \br{1}{B_{s+1}}& & \br{1}{B_l} & & \br{2}{R}
}
\]
The algorithm ends when the size of $B$ reaches $1$.
The number $k_t$ for $k_t$-chunk-sort increases rather fast according to $k_{t+1} = 2^{k_t}$ (line \ref{kGrowthLine})
starting with $k_1 = 2$. On the first epoch $L$ and $R$ are empty.

\begin{lemma}\label{mnEpochLemma}
Each epoch of Algorithm~\ref{logstarAlg} takes $\mathcal{O}(mn)$ time.
\end{lemma}
\begin{proof}
Indeed, calls to ``delete'' take $\mathcal{O}(mn)$ time in total for the reasons 
discussed in the proof of Theorem~\ref{mnlogmnTheorem}.
At each epoch $t$ a $k_t$-chunk-sort of a subarray of length $l_t$ is performed,
which takes $\mathcal{O}(l_t \log k_t)$ time.
The next epoch $t+1$ deals with a subarray of size $l_{t+1} = l_t/k_t$ and
performs a $k_{t+1}$-chunk-sort, where $k_{t+1} = 2^{k_t}$ and therefore, 
takes $\mathcal{O}(l_{t+1} \log k_{t+1}) = \mathcal{O}(\frac{l_t}{k_t} \log 2^{k_t}) = \mathcal{O}(l_t)$ time.
Since $l_t \leq 2mn$ for all $t$, $k_t$-chunk-sort for $t \geq 2$ takes $\mathcal{O}(mn)$ time.
And since $k_1 = 2$, $k_1$-chunk-sort also takes $\mathcal{O}(mn)$ time.
\end{proof}

\begin{theorem}
    Algorithm~\ref{logstarAlg} takes $\mathcal{O}(mn \log^* mn)$ time.
\end{theorem}
\begin{proof}
    Each epoch takes $\mathcal{O}(mn)$ time according to Lemma~\ref{mnEpochLemma} 
    and performs a $k_t$-chunk-sort of an array of size no more than $2mn$.
    When $k_t$ becomes larger than $2mn$, the algorithm obviously stops. 
    Since $k_{t+1} = 2^{k_t}$, and $k_t \leq 2mn$, it takes $\mathcal{O}(\log^* mn)$ epochs to complete the algorithm.
\end{proof}

\section{Conclusions}
We have presented an algorithm that finds the closed discrete Frechet distance in $\mathcal{O}(mn \log^* mn)$ time,
where $\log^*$ is the iterated logarithm. 

Iterated logarithm is an extremely slow-growing function. 
For practical values of $m$ and $n$ the presented algorithm may not be the fastest.
In this case one could easily modify the algorithm to work only in two epochs.
The first epoch would perform a $(\log mn)$-chunk-sort and the second epoch an {$(mn/\log mn)$-chunk-sort}.
This approach would take $\mathcal{O}(mn\log \log mn)$ time but may be faster in practice.

We would like to note that the technique could be applied to other problems that are solved with binary search
and is not limited to Frechet distance in particular or to computational geometry in general.

\printbibliography
\end{document}